\documentclass[12pt]{article}
\usepackage{graphics}
\input epsf

\newcommand{\be}{\begin{eqnarray}}
\newcommand{\ben}{\begin{eqnarray}\nonumber}
\newcommand{\ee}{\end{eqnarray}}

\renewcommand{\theequation}{\arabic{section}.\arabic{equation}}

\begin{document}
\title{
%\begin{flushright}
%{\large UAHEP041}
%\end{flushright}
%\vskip 1cm
Covalent Molecular Binding in a Susy Background}
\author{L. Clavelli\footnote{lclavell@bama.ua.edu} \quad and Sanjoy K. Sarker\footnote{ssarker@bama.ua.edu}\\
Department of Physics and Astronomy\\
University of Alabama\\
Tuscaloosa AL 35487\\ }
%\today
%\date{June 22, 2006}
\maketitle
\begin{abstract}
     The Pauli Exclusion principle plays an essential role in the structure of
the current universe.  However, in an exactly supersymmetric (susy) universe, the
degeneracy of bosons and fermions plus the ability of fermions to convert
in pairs to bosons implies that the effects of the Pauli principle would be
largely absent.  Such a universe may eventually occur through vacuum decay
from our current positive vacuum energy universe to the zero vacuum
energy universe of exact susy.  It has been shown that in such a susy 
universe ionic molecular binding does exist but homonuclear diatomic
molecules are left unbound.  In this paper we provide a first look at covalent 
binding in a susy background and compare the properties of the homonuclear
bound states with those of the corresponding molecules in our universe.
We find that covalent binding of diatomic molecules is very strong in an
exact susy universe and the interatomic distances are in general much smaller
than in the broken susy universe.
\end{abstract}

PACS: 71.30 +h

\section{\bf Introduction}
\setcounter{equation}{0}
\par

    String theory suggests that our current universe with a positive vacuum
energy will in the future make transitions to other minima of the effective
potential.  For a review see ref.\,\cite{Kumar} .This prediction acquires added cogency from the essentially established fact that such a 
transition was made from an inflationary era in the very early universe
to our current universe with a small but non-zero cosmological constant.
Among the possibly numerous local minima of string theory \cite{Giddings}
are some with
exact susy and zero vacuum energy such as in the five original superstrings.
If the universe falls into one of these minima it might remain there 
indefinitely since there are no vacuum fluctuations in these vacua.
It is therefore of potential physics interest to contemplate the properties 
of bulk susy matter.  We envision a scenario similar to that of chain inflation
\cite{Freese} in which a susy ground state of the universe with zero vacuum
energy seems plausible.  However, as exemplified in this paper, in bulk matter
the energy advantage of a system of bosons relative to a system of fermions
with similar interactions far outweighs the vacuum energy advantage.
Even apart from the possibility of a phase transition to exact susy, the 
calculation described in this paper 
could serve in an academic way to elucidate the significance of the
Pauli principle in our universe.
\par
   
     We assume that our current universe has a broken susy with masses of the
bosonic quarks and leptons at the hundred GeV to one TeV scale. 
This may be confirmed in the near future at the Large
Hadron Collider (LHC).
Since, as in Lagrangian Higgs models, the number of degrees of freedom
is preserved in a transition between local minima, we would expect that
any future susy universe would have degenerate fermionic and scalar electrons,
neutrinos, quarks, nucleons etc.  In graphs and tables we assume
for definiteness that the common mass of these
susy multiplets is equal to that of the corresponding standard model particles 
in our broken susy universe where the electron has mass $m=0.511$ MeV.
Other assumptions for the common mass could be investigated
since we give simple expressions for the Hamiltonian expectation
value of the homonuclear diatomic molecules as a function of the common
electron/selectron mass in a simple variational calculation with two
parameters.  

     In the following section II we present this calculation.
We use a generalization of the Heitler-London (HL) wave function 
\cite{HL} first employed 
in the case of the hydrogen
molecule, $H_2$.  A generalization to heavier
elements is complicated in our universe due to the Pauli principle.
However, in a susy universe where electrons freely convert in pairs
to bosonic electrons,  the generalization of the HL wave function
could provide comparable accuracy for the binding of the heavier
elements of moderate atomic number as the original calculation gave 
for $H_2$.  In the summary and discussion presented
in section III we discuss several obvious improvements of the variational
wave function which could be made to improve the accuracy of the
binding energies.  At present, however, it is not our intention to
perform a state-of-the-art molecular physics calculation but to 
present some simple results which are unlikely to be 
qualitatively invalidated by later refinements. 

Certain integrals which occur in the calculation are given analytically 
in appendix A and a method for calculating combinatoric factors is
described in appendix B.  
 

\section{\bf Variational Wave Function for Diatomic Molecules}
\setcounter{equation}{0}

For a multi-electron system of $N$ electrons in the presence of a 
nucleus of charge $Z$ 
the electronic Hamiltonian, including electron-electron repulsion, is
\be
\label{H1}
H=-\frac{1}{2m}\sum_{i=1}^{N} \nabla_i^2-Ze^2\sum_{i=1}^{N} \frac{1}{r_i}+e^2 \sum_{i
< j}^{N} \frac{1}{\left|\vec{r}_i-\vec{r}_j\right|}\quad .
\ee

In a recent paper \cite{CL} a variational wave function for a susy atomic ion of
atomic number $Z$ with $N$ susy electrons was given in the form:
\be  
\psi = \prod_{i=1}^{N} u(\vec{r}_i)
\label{wavefn}
\ee
with
\be
     u(\vec{r}_i)= \sqrt{\frac{Z_s^3}{\pi}} e^{-Z_s |\vec{r}_i|} \quad .
\label{u}
\ee
     
This corresponds to putting all $N$ electrons into a $1s$ wave function
which is a poor approximation for $N>2$ in our world due to the Pauli principle.  However,
in a susy world it could be a quite adequate approximation to the ground state
wave function since fermionic electrons in excited states would convert in
pairs to bosonic electrons (selectrons) which would then drop into the $1s$
ground state via photon emission.  To account for the mutual repulsion between 
electrons,  $Z_s$ in eq.\,\ref{u} is treated as a
variational parameter which minimizes the expectation value of the Hamiltonian
by taking the value
\be
    Z_s = me^2 (Z- \frac{5}{16}(N-1)) \quad .
\label{Zs}
\ee
Therefore the effective nuclear charge seen by an electron is reduced from $eZ$ to 
$eZ_{eff} = e(Z - (N-1)5/16)$ due to screening by other electrons. The radius of the 
atom is correspondingly increased by a factor of $Z/Z_{eff}$.
In treating the covalent binding of two such atoms we will use the wave functions 
of eq.\,\ref{u} centered on two nuclei at $\pm \vec{R}$.  We shall see that the 
value of $Z_s$ that
minimizes the energy will differ slightly from eq.\,\ref{Zs} 

The Bohr radius for hydrogen is
\be
   \frac{1}{me^2} =  a_0 = 0.529 A^\circ  \quad .
\ee
We work in a system of units where $\hbar=c=1$ so 
\be
    2 R_\infty = \frac{1}{m a_0^2} = m e^4
\ee
$R_\infty$ being the Rydberg constant, $13.6$ eV, and $e^2$ being
the fine structure constant $1/137$.
  
The variational estimate of the atomic energy is then
\be
    E(Z,N) = - R_\infty N Z_{eff}^2  = - R_\infty N (Z - 5(N-1)/16)^2 \quad .
\label{Eatomic}
\ee
For the neutral atom ($N =  Z$), $Z_{eff} > 11 Z/16 \propto Z$. Hence with 
increasing $Z$, the atoms become smaller as the radius decreases as $\sim 1/Z$ 
and the binding energy rapidly increases as $Z^2$.   As discussed in section III, 
further refinements of the variational wave function are not expected to change
these results qualitatively.
From Eq. (2.7) we see that energy involved in subtracting or adding electrons to a 
neutral atom increases rapidly with $Z$. For example, the ionization energy of 
$K (Z = 19)$ is $105$ eV, and the electron affinity of $Cl (Z = 17)$ is $146$ eV.  
In contrast, the corresponding energies in our universe are
of the order a few eVs, and do not depend strongly $Z$. 
 
From \ref{Eatomic} it is a simple matter to estimate the classical ionic binding energy
by bringing two such atoms to a distance of twice the atomic radius, taking
$l$ electrons from one atom and adding them to the other taking into account
the consequent Coulomb attraction.  In general the ionic
binding is much greater in the susy system than in the standard model molecules. 
We find that the binding energy increases rapidly with increasing $\Delta Z = |Z_1 -
Z_2|$.  This is because it is energetically favorable for a low-Z atom to give up an 
electron or two to a high-Z atom since ionization energy and electron affinities 
increase rapidly with $Z$.  However, the ionic binding energy is found to be negative 
(anti-bonding) for homonuclear molecules ($\Delta Z = 0$) in both the susy and 
standard model systems.

Covalent bonding is an intrinsically quantum mechanical effect where one or more electrons are
simultaneously shared by two nuclei.  
In a susy system the lowest energy will be
found by the totally symmetric wave function with at most two fermionic electrons.

Our approach is to follow the early quantum mechanical treatments
of molecular bonding while discounting the effects of the Pauli
principle.  We consider a diatomic system of two atoms
each having nuclear charge $Z$ and $N=Z$ electrons. 

One might try the many electron generalization of
the molecular orbital method \cite{Hund} corresponding to the wave function
\be
    \Psi \propto \prod_{i=1}^{2N} \left( u(\vec{r}_i + \vec{R}) + u(\vec{r}_i - \vec{R})\right)
\quad .
\label{PsiOld}
\ee
Here $2 R = D$ is the interatomic distance to be determined variationally. 
Although this gives a rough approximation to the $H_2$ molecule
it would not be a serious candidate for a many-electron molecule
in our universe due to the Pauli principle.  In the susy case
it becomes of interest to consider this wave function for larger
$N$.  The wave function \ref{PsiOld} will be an exact solution in the absence
of e-e interaction. However, for small $\Delta Z$ it does not work well since
the wave function, when multiplied out, contains a preponderance of ionic 
configurations in which one atom has more electrons than the other. 
These are not favorable energetically because of  
electron-electron repulsion within the atom, and as noted above, lead to
unstable ionic bonding for homonuclear molecules.  Further, since the weights
of these configurations do not depend on $R$, the total energy in the wave function 
of eq.\,\ref{PsiOld} does not approach twice
the energy of the isolated atoms given by eq.\,\ref{Eatomic}. In the actual state 
the electrons'
motion will be correlated to keep them apart.  
Indeed, for moderate values of $Z$, the energy expectation
value for the wave function eq\,\ref{PsiOld} is found to be greater than that from the 
generalization given below of the Heitler-London wave function \cite{HL} which therefore provides a better approximation to the true wave function.

We consider, therefore, a variational wave function of the form:  
\be
    \Psi = N_0(N) \left(\prod_{i=1}^{N} u(\vec{r}_i-\vec{R})\right)
       \left(\prod_{j=N+1}^{2N} u(\vec{r}_j+\vec{R})\right) + \displaystyle{permutations}
    (\vec{r}_i \leftrightarrow \vec{r}_j)  \quad . 
\label{PsiHL}
\ee
Here $N_0(N)$ is a normalization constant. Note that there are equal number of electrons
centered on each atom. The permutations make the wave function totally symmetric 
as appropriate to bosonic electrons.  If there are (up to two) fermionic electrons in
the system, they would also be expected to have a ground state in the totally
symmetric state since other possibilities would be partially
anti-bonding.  We do not discuss excited states in the current paper nor do we
consider possible mixing effects between states of differing number of fermionic
electrons.
In the
presence of the other atom, $Z_s$ would be $R$ dependent and would be expected to 
approach that of the isolated atoms given by eq.\,\ref{Zs} as $R\rightarrow \infty$.  
We treat $Z_s$ and $R$ as variational parameters to minimize the energy.

In the general case, the symmetrization of the wave function leads to a complex combinatorial 
problem involving a large number of terms each of which is a product of
$2N$ $u$'s. These can be grouped into configurations with $N + m$ electrons
in the wave function, $u_{} = u(|\vec{r}-\vec{R}|)$, centered on the nucleus at $\vec{R}$ 
and $N-m$ electrons centered at $-\vec{R}$. The solution of this
problem is given in Appendix B. Here we specialize to $m = 0$ which corresponds
to the wave function of eq.\,\ref{PsiHL}. In this case there are 
$(2N)!/N!^2$ permutations, including the stated reference
configuration where the first $N$ electrons are 
centered at $\vec{R}$ and the second $N$ electrons are centered at 
$-\vec{R}$. Among the permutations, the number of terms with $k$ 
interchanges is the square of the binomial coefficient 
\be
    n(k) = \left(\begin{array}{c}N\\k\end{array}\right)^2\quad . 
\ee
Note that 
\be
    \sum_{k=0}^{N} n(k) = (2N)!/N!^2 .
\ee
As a function of $Z_s$ and $R$ we minimize the expectation
value of the Hamiltonian written as a sum of kinetic, electron
-nucleus Coulomb terms, electron-electron correlation terms,
and nucleus-nucleus repulsion term.
\be
   H_{kin}&=& - \frac{1}{2 m} \sum_{i=1}^{2N} \nabla_{i}^2\\
   H_{Ze}&=& - Ze^2 \sum_{i=1}^{2N}\left(\frac{1}{|\vec{r}_i
-\vec{R}|}+\frac{1}{|\vec{r}_i+\vec{R}|}\right)\\
   H_{ee}&=& e^2 \sum_{i<j}^{2N}\frac{1}{|\vec{r}_i-\vec{r}_j|}\\
   H_{nuc}&=& Z^2 e^2/|2\vec{R}|
\ee
The corresponding expectation values are written in terms of
one-body and two-body integrals, $A_i$, summarized in appendix A
together with combinatoric factors discussed in appendix B.
Each expectation value takes the form of a sum over $k$, the
number of interchanges. The normalization of the wave function, for
example, requires that
\be
      \left(N_{0}(N)\right)^2 \left(\begin{array}{c}2N\\N\end{array}\right)=
\left( \sum_{k=0}^{N} n(k) A_0^{2k} \right)^{-1}
\label{normalization}
\ee
The results are
\be
    <\Psi|H_{kin}|\Psi> = \left(N_0(N)\right)^2 \left(\begin{array}{c}2N\\N\end{array}\right)
\frac{Z_s^2}{m} \sum_{k=0}^{N} n(k)\left((N-k)A_0^{2k} + k A_1 A_0^{2k-1}\right)
\ee 
\ben
    <\Psi|H_{Ze}|\Psi> = - 2 \left(N_0(N)\right)^2 \left(\begin{array}{c}2N\\N\end{array}\right)
Z_s Z e^2 \sum_{k=0}^{N} n(k)\left((N-k)(1+A_{2a})A_0^{2k} + k A_{2b} A_0^{2k-1}\right)\\ 
\ee 
\ben
   <\Psi|H_{ee}|\Psi> &=& \left(N_0(N)\right)^2 \left(\begin{array}{c}2N\\N\end{array}\right)Z_s e^2 
\sum_{k=0}^{N} n(k)\left( A_{3c} A_0^{2k-2} k(2k-1) 
+ 4k(N-k)A_{3b} A_0^{2k-1}\right.\\ 
&+& \left. A_0^{2k}(N-k)(I_3 (N-k-1) + A_{3a}(N-k))\right) \quad . 
\ee 
In the large $R$ limit 
\be
  <\Psi|H|\Psi> \rightarrow N\left( Z_s^2/m - 2 e^2 Z_s (Z - 5 (N-1)/16)\right)
\ee
This is a minimum at
\be
      Z_s = m e^2 (Z - 5(N-1)/16) = (Z-5(N-1)/16)/a_0
\ee
with
\be
      <H> \rightarrow - 2 R_\infty N(Z - 5(N-1)/16)^2 \quad .
\ee 
Thus at large separation the energy of the diatomic system approaches twice the energy
of each isolated atom given in eq.\,\ref{Eatomic}.  By examining the asymptotic forms
for the integrals $A_i$ one can see that for large $R$ the asymptotic energy is
generally approached exponentially.

At intermediate values of the separation we write
\be
     Z_s = (Z - 5(N-1)/16)(1+\delta)/a_0 \quad .
\ee

We vary $\delta$ and the interatomic separation $2 R$ to minimize the energy.
In figure 1 we show for $H_2$ the expectation value $<H>$ as a function of $R$
at the optimum value of $\delta$ for each $R$.  
In table 1 we give the results of the overall minimization for the eight lowest 
homonuclear diatomic molecules.  We show for each molecule the minimum energy
and equilibrium interatomic separation as well as the overall optimum value of 
$\delta$.  We neglect the vibrational energy of the nuclei in the ground state
so we do not distinguish between the minimum value of the Hamiltonian and the
negative of the dissociation energy. 

\begin{figure}[htbp]
\begin{center}
\epsfxsize= 3in %% 6.7in % actual
\leavevmode
\epsfbox{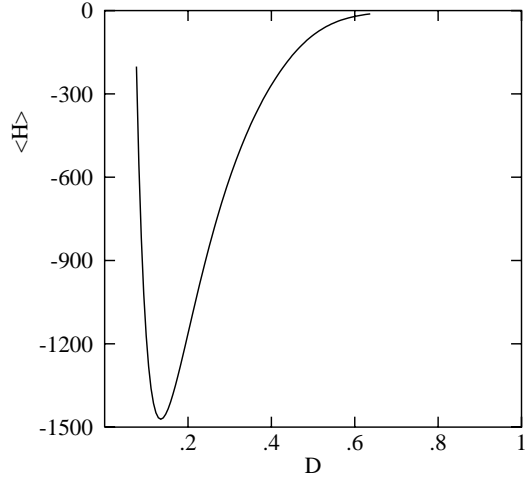}
\end{center}
\caption{For the case of the $O_2$ molecule with bosonic constituents the graph
plots the expectation value of the Hamiltonian in $eV$ as a function of the interatomic
distance $D$ in $A^\circ$.  For each value of $D$, the $\delta$ parameter is chosen to
minimize the energy.}
\label{O2pot}
\end{figure}

\begin{table}[htbp]
\begin{center}
\begin{tabular}{||c|ccc|cc||}\hline
   &                  &                   &             &                          &        \\

   &  $D$ ($A^\circ$) & $H_{min}(susy)$   &   $\delta$  &   $D {\displaystyle{(fermionic)}}$   &   
$B.E. {\displaystyle{(fermionic)}}$  \\
\hline
\hline
       &           &          &         &          &         \\
$H_2$  &    0.748  &  -3.782  &  0.166  &  0.74    &   4.71 \\
$He_2$ &    0.454  &  -27.15  &  0.208  &  2.97    &  0.0009\\
$Li_2$ &    0.326  &  -86.04  &  0.222  &  2.67    &  5.4   \\
$Be_2$ &    0.254  &  -196.3  &  0.228  &    -     &    -    \\
$B_2$  &    0.208  &  -374.0  &  0.231  &  1.59    &  3.0   \\
$C_2$  &    0.176  &  -635.0  &  0.234  &  1.24    &  6.21  \\
$N_2$  &    0.153  &  -995.6  &  0.235  &  1.10    &  9.76  \\
$O_2$  &    0.135  & -1471.6  &  0.237  &  1.21    &  5.12  \\
       &           &          &         &          &         \\
\hline
\end{tabular}
\end{center}
\caption{ 
Binding properties of the eight lightest homonuclear diatomic molecules with totally
symmetric electronic wave function as would be appropriate in the susy case.  In the second 
column we give the interatomic distance ($D = 2R$) in angstroms.  The corresponding minimum energy
expectation value is given in $eV$ in the third column.  The fourth column gives the $\delta$
parameter at the minimum.  The fifth and sixth columns give the experimental \cite{Huber}
interatomic 
distance in angstroms and the experimental binding energy in $eV$ for the diatomic 
molecules of the current universe (fermionic constituents). Neglecting vibrational
energy the experimental binding energy is the negative of the potential at the minimum.}
\label{properties}
\end{table}

\section{\bf Discussion}
\setcounter{equation}{0}

    We have studied the molecular properties of the first eight homonuclear diatomic
molecules in a totally symmetric HL electronic ground state as would be appropriate in a
susy background where fermionic and bosonic electrons are degenerate in mass.
This wave function was used for $H_2$ by Wang \cite{Wang} in the early days of quantum mechanics.

Our results for $H_2$ are in good agreement with those of that paper obtained in the days before computers.
For instance our results from table 1 for the $H_2$ case can be compared with his
interatomic distance of 0.76 $A^\circ$, minimum energy of $-3.76$ eV, and $\delta$
parameter $0.166$.  The $H_2$ calculation, but not that for heavier molecules, is the 
same for bosonic as for fermionic electrons due to the absence of Pauli effects.

Our model calculation and that of Wang is $20 \%$ below the experimental binding
energy of $4.71$ eV.  For the interatomic distance these calculations are good to
about $1\%$.  We can expect that similar accuracies would hold in the susy 
case for the diatomic molecules above $H_2$.  The interatomic distances predicted in
table \ref{properties} are somewhat less than twice the radii of the isolated atoms
$3/(2Z_s)$. 

    In the case of the higher molecules, our calculation shows progressively smaller
interatomic distances compared to the experimental values for the molecules with
fermionic electrons which are all of order of $1 A^\circ$. The third column of table 1 shows the energy of diatomic molecule relative to the total energy of two isolated atoms. We see that molecules are much more strongly bound than the corresponding molecules with fermionic constituents. The binding energy increases rapidly with $Z$. In contrast, in the standard model molecular binding energy
is of the order of few eVs in all cases (column 6). These facts suggest
that molecules in a susy background would have much lower reaction cross sections 
than the diatomic molecules in our broken susy background. 

From the second derivative of the energy at the minimum one can estimate the 
first vibrational energy level.
\be
    \frac{1}{2}\hbar \omega = \frac{\hbar}{2} \sqrt{\frac{d^2<H>}{dR^2}/M}
\ee
where $M$ is the nuclear mass.  From the curve of figure \ref{O2pot} we can estimate
that the vibrational ground state lies, in the case of $O_2$, $4.4$ eV or $0.3\%$ 
above the minimum
of the potential.  This compares with $0.196$ eV in the case of the usual $O_2$ molecule.
\cite{Huber}.

In summary, the tendency of bosonic electrons to accumulate in the low-energy regions leads to strongly bound atoms and molecules in a susy world. In a susy atom 
all electrons occupy the $1s$ orbital in the ground state, leading to a spherical shape whose radius decreases as $1/Z_{eff}$, and the total energy scales as - $Z_{eff}^2$. The important point is that nuclear charge is only partially shielded,
so that $Z_{eff} \propto Z$, which leads to ionization energies and electron  
affinities that increase rapidly with increasing $Z$. Similarly, atoms strongly 
bind to form molecules with binding energies that increase rapidly with $Z$.
For example, a binding energy of 1471 eV for $O_2$ corresponds to a dissociation temperature 
of $ \sim 10^7 K$. These properties have important consequences for bulk matter in the susy universe. It suggests that complex molecules with 
larger number of atoms would form easily and that solids might have high melting 
points. Of course, macroscopic systems would exhibit Bose condensation and superfludity with high critical temperature. 

The situation is very different in our universe, where only valence electrons
take part in chemical reactions and molecular bonding, as they occupy the outermost 
orbitals due to the Pauli principle. The other electrons  
are effectively frozen out as they occupy the inner orbitals; their primary function is
to shield the nuclear charge, which they do quite efficently. Consequenly $Z_{eff} \sim 1$, 
and ionization energies and electron affinities, and molecular binding energies
are all of order a few eVs for all atoms. 

    As in every variational calculation, the accuracy can be improved and the energy
estimate lowered by adding additional variational parameters.  Asymptotically, the exact
ground state energy will be approached from above.
In the case of $H_2$ an improved
energy was obtained \cite{Weinbaum} in the early days by adding a single parameter
corresponding to a term in the wave 
function where both electrons were centered on the same nucleus.  Reference \cite{Weinbaum}
found a value of $4.0$ eV for the dissociation energy compared to the experimental value of
$4.7$ eV.  

For higher molecules similar refinements of the wave function would lead to 
an improvement in energy, but the main results are not expected to change qualitatively. In the case of a single atom, the key issue is the screening of the nuclear charge. Obviously the result can be improved by using the self-consistent Hartree-Fock approximation, in which the one-particle wave function
$u(\vec r)$ in \ref{wavefn} itself is determined variationally.
Because of the spherical symmetry, the screening electric field seen by an electron at $\vec R$ is due to the fraction of the other $N-1$
electrons that are within the sphere of radius $R$. Hence the average
screening charge is given by 
\be 
Q_{scr}(R) = -(N-1)e \int _{0}^R dr 4\pi r^2 u^2(r) \quad . 
\ee
Since the function $u(r)$ is the same for all electrons, 
a finite fraction of each electron is outside the sphere. Only when $R \rightarrow \infty$, does the integral equal unity, and $Q_{scr} = -(N -1) e$. 
On the other hand, suppose we take $R$ to be the most probable distance, obtained by minimizing $r^2u^2(r)$.
Then roughly half the particles will be outside the sphere of radius R, and hence roughly half of $Z$ nuclear charge will be unscreened. This means that, $Z_{eff} \propto Z$.
Note that this argument does not depend on the precise form of $u(r)$. Hence,
our results are qualitatively correct. Similar quantitative improvement
can be achieved by introducing two-body (so called) Jastrow 
factors of the form $\prod _{ij}f(\vec{r}_i - \vec{r}_j)$ in the wave function, which take into account correlation effects by keeping particles away from each other. The problem then becomes intractable for large $Z$. However, we do not expect the qualitative physics to change.

For molecules results can also be improved by including admixtures of ionic 
configurations -- that is, configurations with $Z+m$ particles centered on one atom, and $Z-m$ atoms centered on the other, with the relative amplitude determined variationally. These effects will lower the energy, and thus 
make the molecules even more strongly bound. However, since the transition to exact susy is unlikely to occur in the near cosmological 
future, it does not seem urgent at this time to seek an accuracy beyond that 
obtained by our present approximation.   

{\bf acknowedgements}\\
   This work was supported in part by the US Department of Energy under
grant DE-FG02-96ER-40967.  We acknowledge helpful discussions with R. Tipping.

\appendix
\renewcommand{\theequation}{\Alph{section}.\arabic{equation}}
\section{\bf Overlap Integrals}
\setcounter{equation}{0}

Since the wave function is totally symmetric in the $2N$ 
electron positions, overlap integrals can be written as
the number of permutations $(2N)!/N!^2$ times the overlap
of a reference configuration with the full wave function.
For the reference configuration we take the wave function
where the first $N$ (scalar) electrons are in $1s$ states
concentrated around the nucleus at position $\vec{R}$ while
the remaining electrons are concentrated at $-\vec{R}$.

\be
    \Psi_{\displaystyle{ref}} = N_0(N) \left(\prod_{i=1}^{N} u(\vec{r}_i-\vec{R})\right)
       \left(\prod_{j=N+1}^{2N} u(\vec{r}_j+\vec{R})\right) = |--\ldots -++\ldots +>
\label{Psiref}
\ee
Then the matrix element of any operator takes the form
\be
     <\Psi| {\cal{O}}|\Psi> = N_0(N)^2 \left(\begin{array}{c}2N\\N\end{array}\right)
     <\Psi_{\displaystyle{ref}}|{\cal{O}}|\Psi_{\displaystyle{ref}} + \displaystyle{perms}>
\quad .
\ee
If ${\cal{O}}=1$ we have the normalization condition of eq.\,\ref{normalization} with
\be
 A_0 = <-|+> = \frac{Z_s^3}{\pi} \int d^{3}r e^{-Z_s|\vec{r}-\vec{R}|-Z_s|\vec{r}+\vec{R}|}
          = e^{-2 R^\prime}(1 + 2 R^\prime + 4 {R^\prime}^2/3) 
\ee
where
\be
      R^\prime = Z_s R
\ee
%     u(\vec{r}_i)= \sqrt{\frac{Z_s^3}{\pi}} e^{-Z_s |\vec{r}_i|} \quad .

The Hamiltonian consists of a sum over single particle terms and two particle terms.
Its expectation value is given by a combinatoric sum over single and double particle
expectation values.  For instance
\be
   I_1=  <-|-\nabla^2|->=<+|-\nabla^2|+> = -\frac{1}{\pi}\int d^3r e^{-2|\vec{r}|}(1-2/|\vec{r}|) = 1
\ee
and
\ben
    A_1 &=& <-|-\nabla^2|+> = <+|-\nabla^2|-> = \int d^3r u(\vec{r}-\vec{R})(-\nabla^2)
u(\vec{r}+\vec{R})\\
 &=& e^{-2 R^\prime}(1 + 2 R^\prime - 4 {R^\prime}^2/3)
\ee

The electron-nucleus interaction involves the integrals
\be
    I_2 &=& <-|\frac{1}{|\vec{r}-\vec{R}|}|->  =<+|\frac{1}{|\vec{r}+\vec{R}|}|+> = 1\\
    A_{2a} &=& <-|\frac{1}{|\vec{r}+\vec{R}|}|-> = <+|\frac{1}{|\vec{r}-\vec{R}|}|+>\\
    A_{2b} &=& 2 <-|\frac{1}{|\vec{r}\pm\vec{R}|}|+> = 2 <+|\frac{1}{|\vec{r}\pm\vec{R}|}|->
\ee
It is easy to show that
\be
    A_{2a} &=& -1 +(1+\frac{1}{2R^\prime})(1 - e^{-4R^\prime})\\
    A_{2b} &=& 2 e^{-2R^\prime}(1+2R^\prime)
\ee
The electron-electron repulsion term involves the two body integrals
\be
    I_3 &=& <--|\frac{1}{|\vec{r}_1-\vec{r}_2|}|-->= <++|\frac{1}{|\vec{r}_1-\vec{r}_2|}|++>\\
    A_{3a} &=& <-+|\frac{1}{|\vec{r}_1-\vec{r}_2|}|-+>= <+-|\frac{1}{|\vec{r}_1-\vec{r}_2|}|+->\\
    A_{3b} &=& <-+|\frac{1}{|\vec{r}_1-\vec{r}_2|}|++>= <-+|\frac{1}{|\vec{r}_1-\vec{r}_2|}|-->\\
    A_{3c} &=& <--|\frac{1}{|\vec{r}_1-\vec{r}_2|}|++>= <++|\frac{1}{|\vec{r}_1-\vec{r}_2|}|-->
\ee
All of the integrals of this appendix except for $A_{3b}$ occurred in the analysis of the
$H_2$ molecule in the 1920's and were reviewed in the book of Pauling and Wilson \cite{Pauling}.
We have added the evaluation of $A_{3b}$ since it occurs in the binding of the higher 
diatomic molecules in the susy case.  The analytic expressions are
\be
    I_3 &=& 5/8\\
    A_{3a} &=& \frac{1}{2R^\prime}(1 - e^{-4R^\prime}) - e^{-4R^\prime}(11/8 + 3R^\prime/2
+2{R^\prime}^2/3)\\
    A_{3b} &=& e^{-2R^\prime}\left(\frac{1-e^{-4R^\prime}}{32 R^\prime}(5+4R^\prime) + 2R^\prime\right)
\ee
and
\ben
    A_{3c} &=& \frac{1}{5}e^{-4R^\prime}\left( 25/8 - 23 R^\prime /2 - 12 {R^\prime}^2
- 8 {R^\prime}^3/3 + \frac{6}{2 R^\prime}(1 + 2R^\prime + 4{R^\prime}^2/3)^2 
(\gamma + \ln(2R^\prime))\right.\\\nonumber
  &+& \frac{6}{2 R^\prime}(e^{8R^\prime}\displaystyle{Ei}(-8R^\prime)(1 - 2 R^\prime + 4 {R^\prime}^2/3)^2\\
   &-&\left. 2 e^{4R^\prime}\displaystyle{Ei}(-4R^\prime)(1 - 2 R^\prime + 4{R^\prime}^2/3)(1+2R^\prime+4{R^\prime}^2/3))\right)
\label{Ei}
\ee
The exponential integral function Ei(x) goes asymptotically as $e^{x}$ which makes it convenient
to factor out the overall $e^{-4R^\prime}$ in eq.\,\ref{Ei}.  The complete expression for $A_{3c}$
was given in 1927 by Sugiura \cite{Sugiura}.

\section{\bf Combinatorics}
\setcounter{equation}{0}
%\noindent {APPENDIX B: Combinatorics}  

We consider a diatomic molecule with the two atoms centered at $\vec{R}$ and $-\vec{R}$, 
each with nuclear charge $Z$, with a total of $2 N=2 Z$ electrons. However here, 
for greater generality, we take $Z$ and $N$ to be independent.  The many-body wave function $\Psi$ 
is constructed from single-paricle functions $u(\vec r \pm \vec R)$ centered at $-\vec{R}$ and $\vec{R}$, 
respectively. The function $u(\vec r)$ is similar, but not identical, to the atomic
wave function, as it is calculated variationally. 

The total wave function $\Psi\,$ is a linear combination of functions $\psi _m$ 
having $N + m$ particles centered at $-\vec{R}$ and $N - m$ particles centered at $\vec{R}$,
with $-N \leq m \leq N$:

\be 
\psi_m =  \prod _{i = 1}^{N-m}u(\vec r_i - \vec R)\prod _{j = N-m+1}^{2 N}u(\vec r_j + \vec R)
  + {\displaystyle permutations} \quad . 
\ee

In calculating the expectation values we need to solve a combinatorial problem. This can be done in a compact 
way by introducing a pseudospin variable $\mu$ for each particle such that $\mu = 1$ or $ - 1$ if the particle is centered at $-\vec{R}$ and $\vec{R}$, respectively. Then $\sum _{i=1}^{2N}\mu _i = 2m$. 

Let $u_{\mu}(\vec r) = u(\vec r + \mu \vec R)$.

Then the ground-state wave function is given by
\be 
\Psi = {2N \choose N}^{-1/2} \sum _{m = -N}^N a_m \sum _{\lbrace {\mu }\rbrace} \delta _{m,\sum _j\mu _j/2} \prod _{i = 1}^{2N} u_{\mu _i}(\vec r_i) \quad .  
\label{gswf}
\ee

Where the $a_m$'s are variational parameters. By symmetry $a_m = a_{-m}$. 

Consider first the evaluation of the normalization constant $<\Psi |\Psi>$ which has $2^{4N}$ terms, each
of which is an integral over the $2N$ coordinates, and thus consists of $2N$ factors. 
We assume that each function $u_{\mu}(\vec r)$ is normalized. But, in general, $u_{\mu}$ and $u_{-\mu}$ has a non-zero overlap $A_0$, so that
\be 
\int d^3r u_{\mu}(\vec r)u_{\mu ^{\prime}}(\vec r) = \delta _{\mu,\mu ^{\prime}} + A_0 \delta _{\mu,-\mu^{\prime}} \quad .
\ee

Then the normalization constant has the general form
\be
<\Psi|\Psi> = \sum _{p = 0}^{2N} Z_pA_0^p 
\ee
where $Z_p$ is a combinatorial factor. 
To determine $Z_p$ consider orthonormal functions
\be
v_n(q) = e^{2\pi inq}
\ee  
where $0 \leq q \leq 1$, and $n$ is an integer.
Since 
\be
<v_m|v_n> = \int_{o}^{1} dq e^{2\pi i(n-m)q} = \delta_{n,m} \quad ,
\ee
we can represent the Kronecker
$\delta$ in the wave function \ref{gswf} in a similar manner, leading to

\be 
\Psi = {2N \choose N}^{-1/2} \sum _{m = -N}^N a_m \sum _{\lbrace {\mu }\rbrace} \int_{0}^{1} dq e^{-2\pi imq} \prod_{j = 1}^{2N} u_{\mu_j}(\vec r_j)e^{\pi i\mu_jq} \quad .  
\ee
The representation allows us to integrate over $\vec r_i$ and sum over $\mu_i$ for each $i$ separately, which
gives

\ben
<\Psi|\Psi> &=& {2N \choose N}^{-1}\sum _{m,m^{\prime}} a_m a_{m^{\prime}} \int_{0}^{1} dq\int_{0}^{1} dq^{\prime}e^{2\pi i(m^{\prime}q^{\prime}- mq)}\\
&\cdot& \lbrace e^{\pi i(q-q^{\prime})} + e^{-\pi i(q-q^{\prime})} + A_0(e^{\pi i(q-q^{\prime})} + e^{- \pi i(q-q^{\prime})})\rbrace^{2N} \quad .
\ee
Expanding the factor $\lbrace...\rbrace ^{2N}$, and integrating over $q,q^{\prime}$ and after some rearrangements we finally obtain,

\be 
<\Psi|\Psi> = \sum_{m,m^{\prime}} a_ma_{m^\prime}\sum _{p = |m-m^{\prime}|}^{2N - |m+m^{\prime}|}F(N,p,m,m^{\prime})A_0^p \quad . 
\ee
where,

\be 
F(N,p,m,m^{\prime})=\frac{N!N!\lbrace 1 + (-1)^{(p+m+m^{\prime})}\rbrace}{2 G}
\ee
with 
\ben
G =\lbrace N+(m+m^{\prime}-p)/2\rbrace!
\lbrace N-(m+m^{\prime}+p)/2\rbrace!\lbrace (p+m-m^{\prime})/2\rbrace!
\lbrace(p+m^{\prime}-m)/2\rbrace! \quad . 
\ee
Thus $F$ is zero unless $p+m+m^\prime$ is even.

{\bf Ground-State Energy:}
To evaluate the energy $<\Psi|H|\Psi>$, let us write $H = H_1 + V_{ee}$, where $V_{ee}$ describes the electron-electron interaction, and 

$H_1 = \sum _{i=1}^{2N} H_{i}$ is the sum of one-particle Hamiltonians $H_i$. We first consider $H_1$ whose average is given by $<\Psi|H_1|\Psi> = 2N <\Psi|H_i|\Psi>$ since the
particles are identical. The evaluation of $<\Psi|H_i|\Psi>$ is similar to that of $<\Psi|\Psi>$ since
the integration over the coordinates factorizes. 
Let 
\be
\int d^3r_i u_{\mu}(\vec{r}_i)H_i u_{\mu ^{\prime}}(\vec r_i) = A_1 \delta _{\mu\mu^{\prime}} + A_2 \delta _{\mu,-\mu^{\prime}} \quad .
\ee
Proceeding as before, we obtain, 
\be 
<\Psi|H_1|\Psi> = \sum _{m,m^{\prime}} a_ma_{m^{\prime}}I(m,m^{\prime}), 
\ee
where
\be I(m,m^{\prime}) = \sum _{p = |m-m^{\prime}|}^{2N - |m+m^{\prime}|}F(N,p,m,m^{\prime})\lbrace (2N-p)A_1A_0^p + pA_2A_0^{p-1}\rbrace \quad . 
\ee
Then the expectation value of the electron-electron term is given by
\be
<\Psi|V_{e,e}|\Psi> = \frac{2N(2N-1)}{2} <\Psi|\frac{e^2}{|\vec r_i - \vec r_j|}|\Psi> \quad .
\ee
since all pairs yield the same expectation value. The procedure is similar, except that averaging
over the pair interaction involves integrating over two coordinates.  
Let 
\be 
B_1 &=& \int d^3r_1 d^3r_2 u_{\mu}^2(\vec r_1)\frac{e^2}{|\vec r_i - \vec r_j|}u_{\mu}^2(\vec r_2) \quad .\\
  B_2 &=&\int d^3r_1d^3r_2 u_{\mu}^2(\vec r_1)\frac{e^2}{|\vec r_i - \vec r_j|}u_{-\mu}^2(\vec r_2) \quad .\\
B_3 &=& \int d^3r_1d^3r_2 u_{\mu}^2(\vec r_1)\frac{e^2}{|\vec r_i - \vec r_j|}u_{\mu}(\vec r_2)u_{-\mu}(\vec r_2) \quad .\\
B_4 &=& \int d^3r_1d^3r_2 u_{\mu}(\vec r_1)u_{-\mu}(\vec r_1)\frac{e^2}{|\vec r_i - \vec r_j|}u_{\mu}(\vec r_2)u_{-\mu}(\vec r_2) \quad.
\ee
Then we have

\be 
<\Psi|V_{e,e}|\Psi> = V_1 + V_2 + V_3 + V_4, 
\ee
where
\be 
V_1 &=& \sum _{p = |m-m^{\prime}|}^{2N - |m+m^{\prime}|}F(N,p,m,m^{\prime})\lbrace  (2N-p)(2N-p-2)+ (m+m^{\prime})^2\rbrace B_1A_0^p/4 \quad. \\ 
V_2 &=& \sum _{p = |m-m^{\prime}|}^{2N - |m+m^{\prime}|}F(N,p,m,m^{\prime})\lbrace  (2N-p)^2 - (m+m^{\prime})^2\rbrace B_2A_0^p/4 \quad.\\ 
V_3 &=& \sum _{p = |m-m^{\prime}|}^{2N - |m+m^{\prime}|}F(N,p,m,m^{\prime})(2N-p)p B_3A_0^{p-1}\quad . \\\nonumber
V_4 &=& \sum _{p = |m-m^{\prime}|}^{2N - |m+m^{\prime}|}F(N,p,m,m^{\prime})\frac{B_4}{4}\lbrace\left( p(p-2)+(m-m^{\prime})^2\right) A_0^{p-2}\\
&& \hspace{2in} + \left( (2N-p)^2 - (m+m^{\prime})^2\right) A_0^p\rbrace \quad .
\ee

{\bf Heitler-London limit:} In this case we keep only $m = m^{\prime} = 0$, so that each atom contains exactly $N$ electrons. As $R \rightarrow \infty$, the molecule dissociates into two isolated atoms in their respective ground states. We can set $a_0 = 1$.

Then we have
\be 
<\Psi|\Psi> = \sum _{k=0}^{N} {N\choose k}^2 A_0^{2k}. 
\ee
\be <\Psi|H_1|\Psi> = \sum _{k=0}^{N} {N\choose k}^2 \lbrace 2(N-k)A_0^{2k}A_1 + 2kA_0^{2k-1}A_2\rbrace. 
\ee
\ben 
<\Psi|V_{e,e}|\Psi> &=& \sum _{k=0}^N {N \choose k}^2\left( A_0^{2k}\lbrace (N-k)(N-k-1)B_1 + (N-k)^2(B_2 + B_4)\rbrace\right.\\
 &+& \left. 4k(N-k)A_0^{2k-1}B_3 + k(k-1)B_4A_0^{2k-2}\right). 
\ee

If we include $m,m^{\prime} = \pm 1$, there is one extra variational parameter $a_1 = a_{-1}$.
These contribute the following terms to $<\Psi|\Psi>$:
\be 
\sum _{k=0}^{N} {N\choose k}^2 A_0^{2k}\lbrace 4a_1A_0 \frac{N-k}{k+1} + 2a_1^2( \frac{N-k}{N-k+1} + \frac{k}{k+1})\rbrace. 
\ee
This analysis is given for completeness and as a guide to future improvements only. In the current paper, numerical results are presented for the $m=m^\prime=0$ term only.
\pagebreak                                                                      

\end{document}